\documentclass{elsart}

\usepackage{graphicx}
\usepackage{color}
\usepackage{amssymb}
\usepackage{multirow}
\begin{document}
\begin{frontmatter}

\title{Cosmogenic Activation of Materials Used in Rare Event Search Experiments}
\author[usd]{C. Zhang}
\author[usd,yze]{D.-M. Mei\corauthref{cor}}
\corauth[cor]{Corresponding author.}
\ead{Dongming.Mei@usd.edu}
\author[uos]{V. A. Kudryavtsev}
\author[lbl]{S. Fiorucci}
\address[usd]{Department of Physics, The University of South Dakota, Vermillion, SD 57069}
\address[yze]{School of Physics and Optoelectronic, Yangtze University, Jingzhou 434023, China}
\address[uos]{Department of Physics and Astronomy, University od Sheffield, Sheffield S3 7RH, UK}
\address[lbl]{Lawrence Berkeley National Laboratory, 1 Cyclotron Rd., Berkeley CA 94720, USA}
\begin{abstract}
We evaluate the cosmogenic production rates in some materials that are commonly used 
as targets and shielding/supporting components for detecting rare events. 
The results from Geant4 simulations and the calculations of ACTIVIA are compared with the available
experimental data. We demonstrate that the production rates from the Geant4-based simulations agree with the available data
reasonably well. As a result,
 we report that the cosmogenic production of several isotopes in various materials 
can generate potential backgrounds for direct detection of dark matter and neutrinoless double-beta decay. 

\end{abstract}


\end{frontmatter}
\section{Introduction}
Evidence from galactic and extragalactic observations indicate 
the existence of dark matter in our universe
\cite{blum, davis, benn}. 
As a favored dark matter candidate, the Weakly Interacting Massive Particle (WIMP) can be 
directly detected by underground experiments through its elastic scattering
off ordinary target materials~\cite{goodman, feng}. 
None of the current dark matter experiments has convincingly observed WIMP scattering events.  
The next generation ton-scale dark matter experiments, 
especially the xenon-based detectors~\cite{lz, xenonton}
are designed to achieve ultra-low background conditions thus allowing 
detection sensitivity to WIMP-nucleon spin-independent cross section down to
$\sim$$10^{-48}\,\rm{cm}^{2}$. 
In order to achieve such a sensitivity level, the background rate in the 
region of interest needs to be at a maximum level of $\sim$0.1$-$0.2 events/ton-year~\cite{lz}.

Neutrinoless double-beta decay experiments are aimed at measuring the effective Majorana mass 
of the electron neutrino down to 10 to 50 meV to understand the nature
(Dirac or Majorana) of neutrinos~\cite{Ell02, Ell04, Avi04, Bar04}. This requires measurements of
a half-life for a nucleus at a level of $>$10$^{27}$ years. 
Existing experiments have achieved~\cite{gerda, exo, kamcan}
or will soon achieve~\cite{MJ} a sensitivity of the order of 10$^{25}$ years
for several isotopes and set an upper limit on the effective Majorana mass of electron neutrino $<$ 0.2 eV.
Eventually, these future experiments~\cite{gerda, exo, MJ, cuore, nemo} target 
a sensitivity of $>10^{27}$ y or $<$ 1 event/ton-year in the region of interest 
to explore mass values 
favoured by both inverted and normal mass ordering.
\par
 Such low-background event rates for both dark matter and neutrinoless double-beta decay
 require the radioactivity level of every detector
component to be accurately measured.
In addition to natural radioactivity, 
cosmogenic activation can add more radioactivity to a detector component. 
Mitigating measures such as underground storage as early as possible can be employed, but accurate
activation calculations are needed to make decisions on design and operations.
The activation of materials for underground experiments 
has been evaluated using ACTIVIA~\cite{luxback, xenon100back}, TALYS~\cite{mei}, and other tools~\cite{gerda,MJ}.
The discrepancy in the estimated activation rates between different tools exists and this deserves an 
investigation of the sources that may cause the discrepancy.
 
\par
In this paper, we evaluate cosmogenic production of radioactive isotopes at sea level
in various materials used for rare event experiments. 
The activation rates are obtained using Geant4-based simulations~\cite{geant4} and the calculations of ACTIVIA~\cite{activia}. 
The results are compared with some experimental data~\cite{luxback, xenon100back}. 

\section{Evaluation of cosmogenic production of radioactive isotopes on the surface}
\subsection{Evaluation tools and input energy spectra}

The Geant4 (V9.5p02)+Shielding modular physics list~\cite{shielding} is used for this study. It includes a set
of electromagnetic and hadronic physics processes, with modular physics, boson physics, lepton physics,
hadron physics, meson physics, nucleon physics, hyperon physics, antibaryon physics, ion physics, and Quark Gluon string model
($>$ 20 GeV), Fritiof string model ($>$ 5 GeV), Bertini-style Cascade ($<$ 10 GeV), 
as well as high precision neutron model ($<$ 20 MeV), required for 
high energy or underground detector simulations. 
For each physics model,
G4MuonNuclearProcess was activated to simulate muon-nuclear inelastic scattering.  
\par
 The modified Gaisser's formula~\cite{gaisser,guan} (Eq.(2) in Ref.~\cite{guan}) is used to sample muons right above the simulation geometry.
 The energy range spans 1 GeV to 100 TeV. The total flux is normalized to be 0.014 cm$^{-2}$s$^{-1}$, which is the 
total muon flux corresponding to the energy range from 1 GeV to 100 TeV on the surface~\cite{gaisser}. The stopping
muons were not included in the simulation due to a much smaller flux~\cite{jrb, hbi}. Although there is a non-negligible fraction of stopping muons at the surface, the main contribution to the cosmogenic activation comes from atmospheric neutrons so stopping muons can be neglected.  

\begin{figure}
\includegraphics[width = \textwidth]{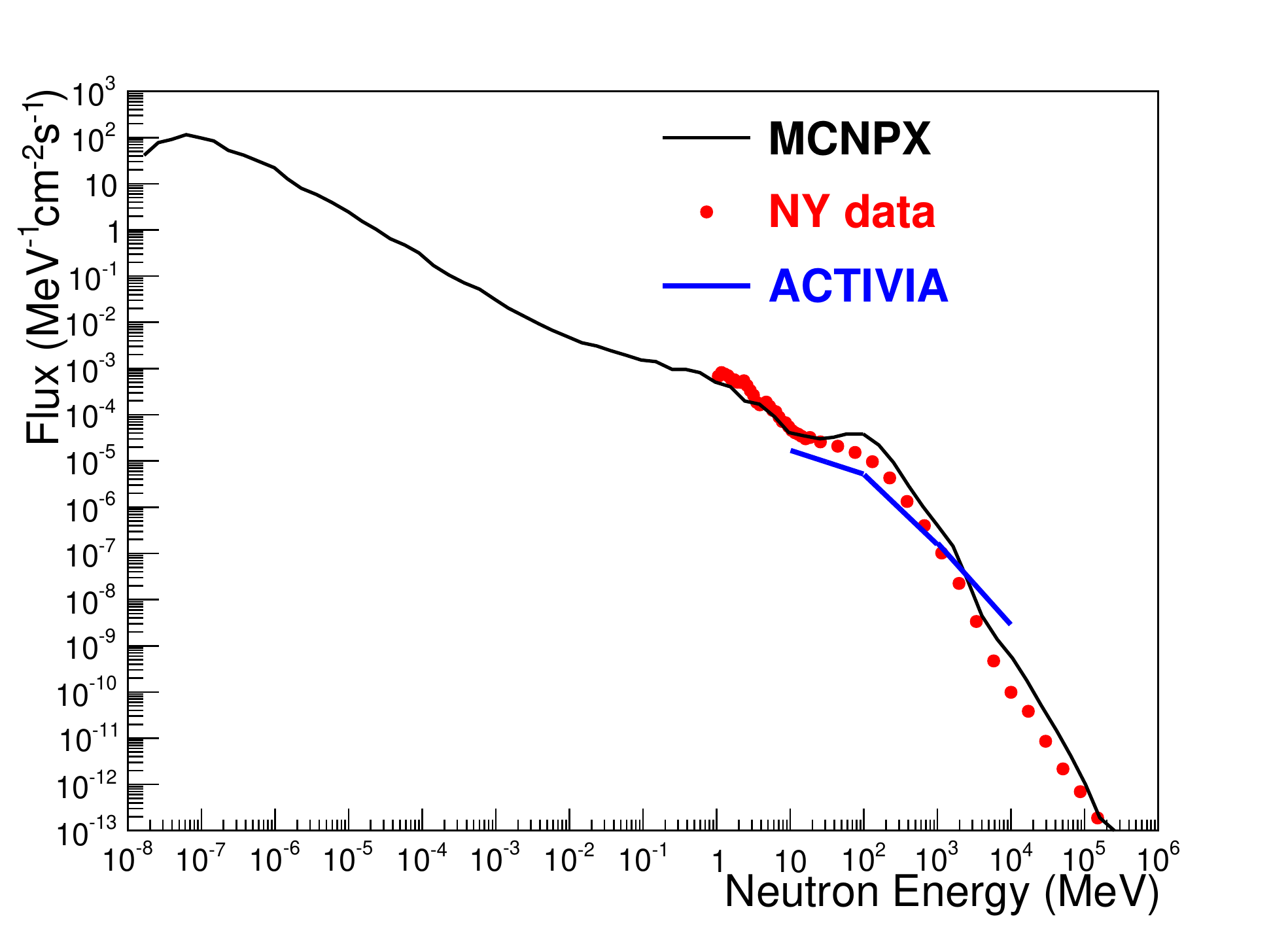}
\caption{\label{NeuSpectrum}
The surface neutron energy spectra adopted in the evaluation. The black curve is the air shower simulation
result using MCNPX package~\cite{chrisLLNL}. The red dots are the measured neutron spectrum at the surface~\cite{gordon}.
The black curve is normalized to the data for $E_{n}>4$ MeV. The blue line is a default input spectrum
used by ACTIVIA package.}
\end{figure}

For surface neutrons, in the Geant4 Monte Carlo simulation, 
the neutron spectrum from thermal neutrons to about 100 GeV neutrons is used. The input energy spectrum of neutrons 
comes from the MCNPX simulation
code~\cite{chrisLLNL} for energies below 4 MeV and the New York data (``NY data'')~\cite{gordon}
 for energies greater than 4 MeV and  is shown in Fig.~\ref{NeuSpectrum}.
The neutron energy spectrum simulated by MCNPX code~\cite{chrisLLNL} is normalized 
to ``NY data'' for neutrons with energy, $E_{n}$, greater than 4 MeV~\cite{gordon}. 
We use normalization to keep the entire spectrum continuous.
This normalization allows us to calculate
the total neutron flux and compare the result to the measured value~\cite{gordon, ldh}. 
Since cosmogenic activation of materials is mainly through neutron capture and neutron inelastic scattering processes and
the latter requires a reaction threshold energy to be usually greater than 4 MeV, we report the output of the activation rates split by neutron energy: (1) thermal neutrons to fast neutrons with energy up to 4 MeV and (2) 
fast neutrons with energy greater than 4 MeV. The energy spectrum for the range of 1$\times$10$^{-8}$ MeV (thermal neutrons) to 4 MeV (fast neutrons) adopts
 the results from MCNPX simulation~\cite{chrisLLNL} and its total flux is normalized to be 0.002 cm$^{-2}$s$^{-1}$,
which is the total neutron flux for this energy range~\cite{chrisLLNL, gordon}. 
For fast neutrons with 
energy greater than 4 MeV, the measurements of ``NY data"~\cite{gordon} is used and its total flux is normalized to 
be 0.004 cm$^{-2}$s$^{-1}$~\cite{gordon}.
It is worth mentioning that the MeV neutron data may have a small 
contamination from local radioactivity, which could slightly overestimate 
the neutron rate from cosmic rays in that range. 
Below the cosmogenic activation rates (atoms kg$^{-1}$day$^{-1}$) from the Geant4 simulation are reported according 
to the production mechanisms - muon spallation, neutron capture ($E_{n}$$<$4 MeV), and neutron inelastic scattering ($E_{n}$$>$4 MeV). 
     
\par
ACTIVIA~\cite{activia} is a package that calculates cosmogenic
production from cosmic-ray activation using data tables and
semi-empirical formulas.
ACTIVIA uses the neutron spectrum at the Earth surface shown 
by the blue curve in Fig.~\ref{NeuSpectrum}. Since this blue curve, the default neutron energy spectrum 
in ACTIVIA, is different from the input energy spectrum used in the Geant4 simulation, we also carried out 
ACTIVIA calculations using the neutron energy spectrum from the ``NY'' data as in the Geant4 simulation,
in order to evaluate any difference between the two energy spectra. 
Therefore, ``ACTIVIA1" stands for ACTIVIA calculation results with the 
blue line as the input neutron energy spectrum. ``ACTIVIA2" represents ACTIVIA results
with the red dots as its input neutron energy spectrum. 

One of our goals is to compare ACTIVIA and Geant4 calculation of activation. 
In addition we compare ACTIVIA calculations of activation using two different neutron energy spectra. 
Since ACTIVIA is widely used in the field of low-background experiments for calculating cosmogenic activation, 
this study may be useful for future evaluation of activation of materials and estimating associated uncertainties. 
The overall difference in the neutron flux is about a factor of 3 
between these two input fast neutron spectra. 

\par
Note that proton activation of materials is also included in the simulation although
the number of protons below a few GeV is less than the number of neutrons in the atmosphere. 
The hadronic part of the cosmic-ray spectrum at the surface is dominated by neutrons~\cite{zei}.
 Proton flux at low energies (below a few GeV) is suppressed because of the proton energy losses. 
Proton spectrum is harder than the neutron one and protons dominate at energies exceeding a few GeV. 
The activation rates of materials 
from protons, with energy spectrum from CRY~\cite{cry}, are provided in Table 1-5.  
It is worth mentioning that ACTIVIA does not include thermal neutron capture.

The variation of muon and neutron fluxes as a function of altitude can be described as~\cite{zei}:
\begin{equation}
\label{altitude}
I_{a} = I_{0}exp(\frac{(A_{0} - A_{a})}{\lambda}),
\end{equation}
where $I_{a}$ is the flux at a given altitude in meters, $I_{0}$ is the flux at the sea level, $A_{0}$ and 
$A_{a}$ are
the atmospheric thickness in g/cm$^{2}$ for the sea level and any given altitudes, $\lambda$ is the 
average absorption length of particles in g/cm$^{2}$ in atmosphere. At the small altitudes, 
typical absorption lengths for muons and neutrons are $\lambda_{\mu}$ = 520 g/cm$^{2}$ and 
$\lambda_{n}$ = 148 g/cm$^{2}$~\cite{zei}. Fig.~\ref{altitude} shows relative intensities
as a function of altitude. As an example, the altitude of the surface laboratory at 
the Sanford Underground Research Facility (SURF) is $\sim$1600
meters above sea level~\cite{luxback}. At this altitude, the muon flux is a factor 1.41,  and the neutron flux is a 
factor 3.34 higher than at sea level. To evaluate cosmogenic activation rates at different altitudes,
the flux correction factors must be taken into account. The production rates are also dependent on
the energy spectrum of muons and neutrons, but in practice the correction of the fluxes is the dominant
effect.  The variation in energy spectra of muons and neutrons is small for altitudes lower than 
2000 meters above sea level. Although there are variations in cosmic radiation flux 
at different latitudes and longitudes, the main contribution to the cosmogenic activation comes from
high-energy neutrons, which do not vary more than 1\% in flux for different 
latitudes and longitudes~\cite{zei}.
\begin{figure}
\includegraphics[width = \textwidth]{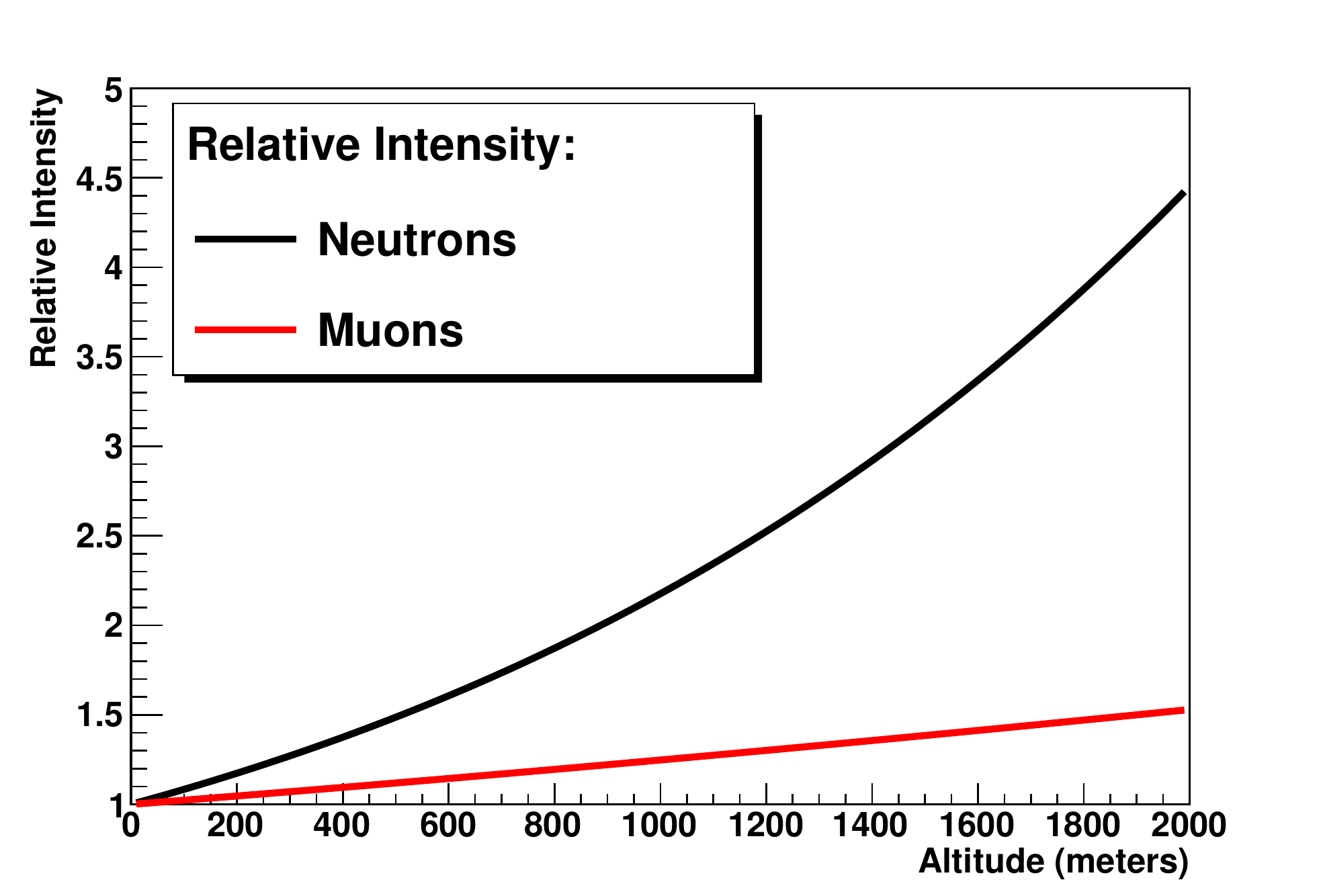}
\caption{\label{altitude} The relative intensity of muons and neutrons as a function 
of altitude. 
}
\end{figure}
\subsection{Cosmogenic production rates in xenon}
Liquid xenon is widely used as a target material 
in dark matter and neutrinoless double-beta decay experiments.
It is generally sealed within gas cylinders made of stainless steel for storage and transportation purposes. 
Although natural xenon is stable, it suffers from cosmic-ray bombardment
 at the Earth surface. The cosmic rays
 produce radioactive xenon isotopes.  The resulting activity can be estimated from the exposure time of 
xenon to cosmic rays
on the surface since the time when xenon was produced.
\par
A comprehensive Geant4 (V9.5p02)~\cite{geant4} simulation has been conducted to evaluate the rate of 
xenon activation. 

The xenon gas is assumed to be stored in a stainless steel cylinder that is 30.48 cm in diameter, 127 cm in height
and 0.762 cm in thickness. 
The pressure of the gas is set to be 970 psi gauge with a density of 0.3657 g/cm$^{3}$ and temperature of 
20 $^{\circ}$C. The total mass of the compressed gas is calculated to be 30 kg per cylinder.
The cosmic-ray muons and neutrons are considered as inputs shot at a xenon gas cylinder. 
The cosmogenic production of xenon isotopes at the surface is dominated by neutron inelastic 
interactions.

Impurities other than radioactive xenon in xenon target can be eliminated by purification. 
However, radioactive xenon isotopes
generated by cosmogenic activations cannot be removed during the purification.  
Low-energy neutrons, typically at thermal energies, activate target materials mainly through 
capture processes. 
The dangerous neutron capture in xenon targets are: $^{126}$Xe($n$,$\gamma$)$^{127}$Xe, $^{132}$Xe($n$,$\gamma$)$^{133}$Xe, 
and $^{134}$Xe($n$,$\gamma$)$^{135}$Xe.

In contrast, high-energy neutrons or muons can break stable xenon nuclei and convert them into radioactive isotopes,
such as $^{127}$Xe, $^{133}$Xe, $^{135}$Xe, $^{125}$I, $^{129}$I, $^{121}$Te, $^{123}$Te, etc. 
Among these production processes, neutron inelastic scattering reactions, $^{128}$Xe($n$,2$n$)$^{127}$Xe, 
$^{129}$Xe($n$,3$n$)$^{127}$Xe,  $^{134}$Xe($n$,2$n$)$^{133}$Xe, $^{136}$Xe($n$,2$n$)$^{135}$Xe, 
dominate the production of radioactive xenon isotopes
as shown in Table~\ref{gasXenon}. 
\par
\begin{figure}
\includegraphics[width = \textwidth]{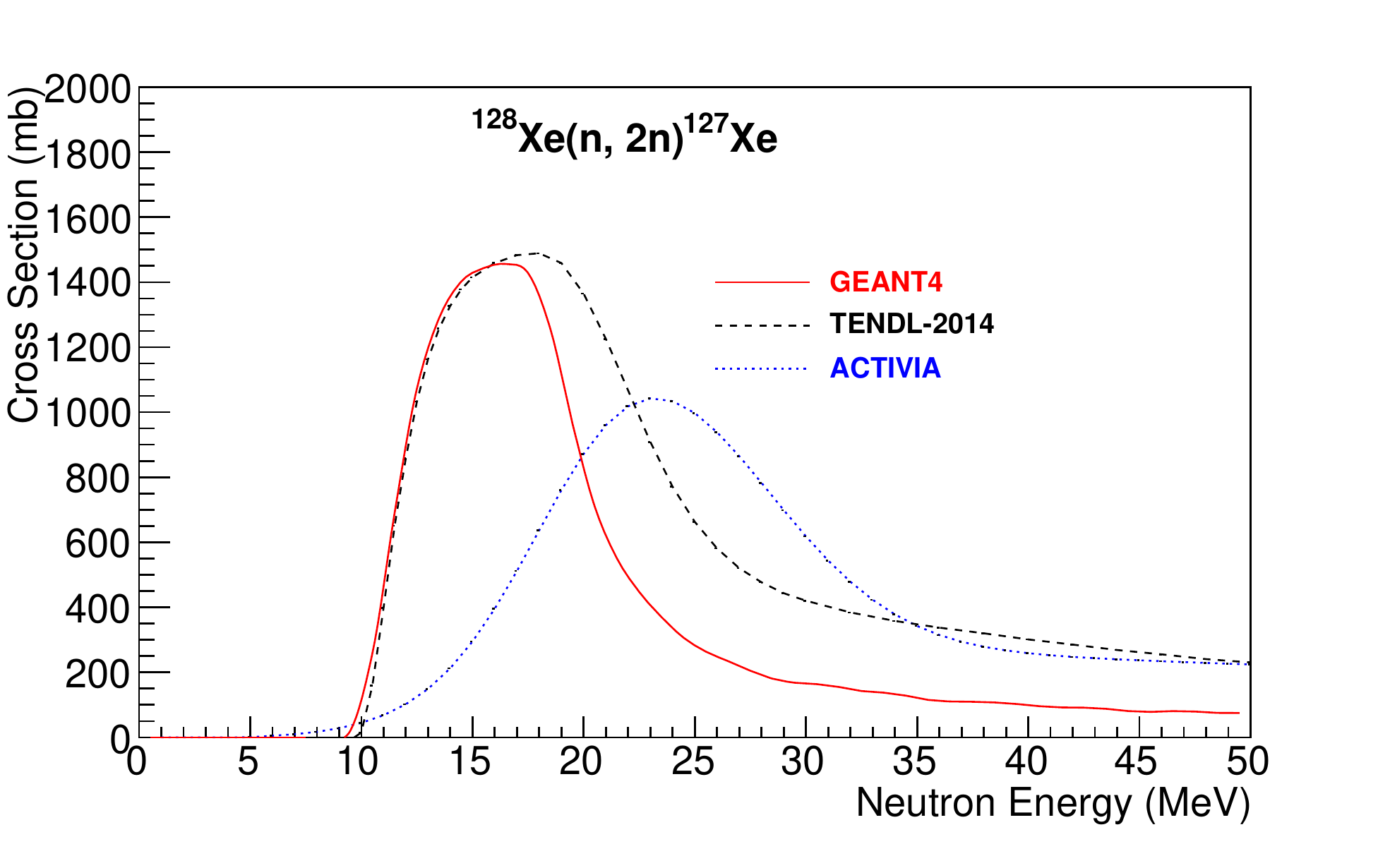}
\caption{\label{xe1271n} Cross-sections comparison for the production of $^{127}$Xe 
through $^{128}$Xe(n,2n)$^{127}$Xe 
between Geant4, ACTIVIA, and TENDL-2014~\cite{arj}, which is a default cross-section library for TALYS-1.8~\cite{arj}.}
\end{figure}
\begin{figure}
\includegraphics[width = \textwidth]{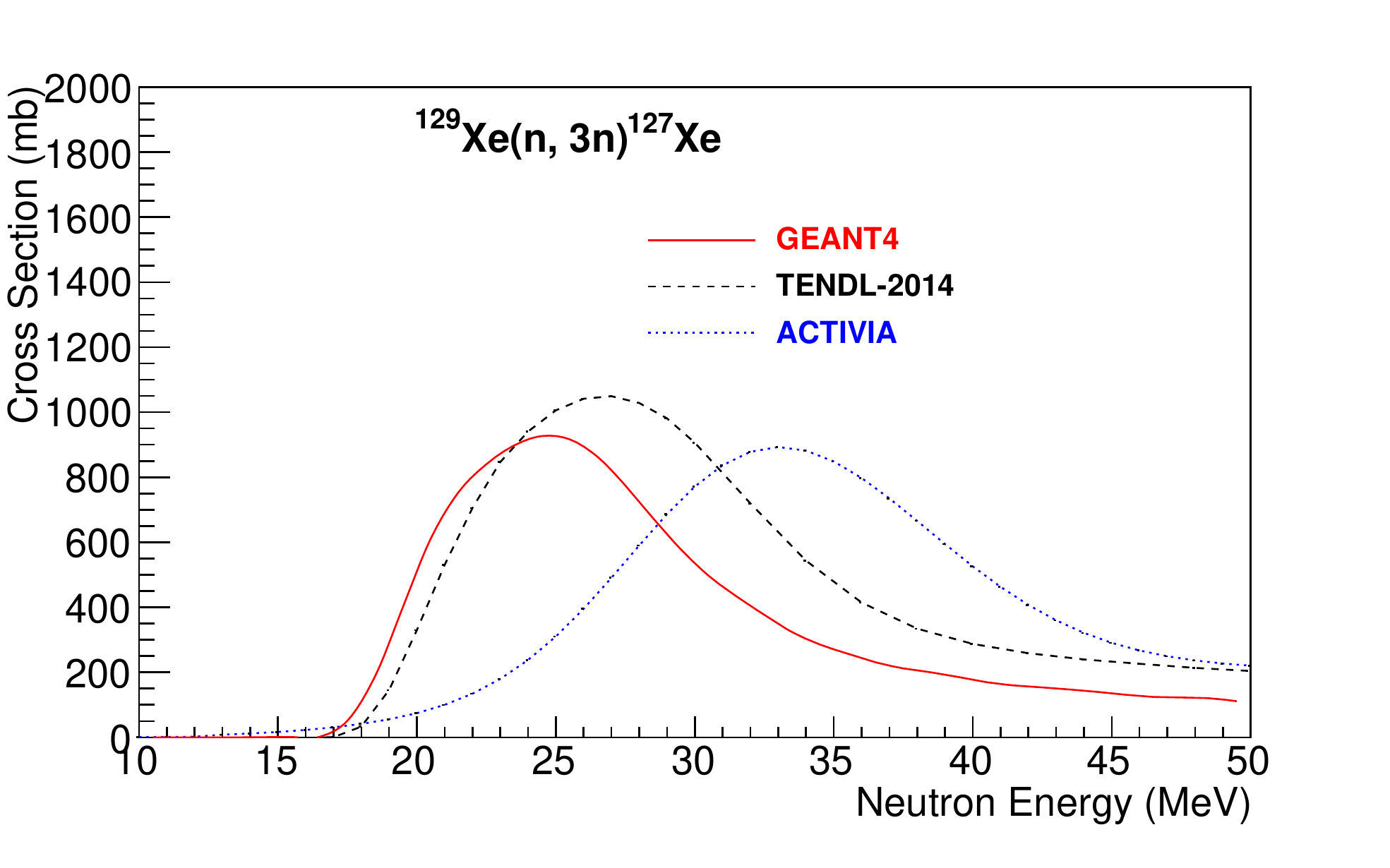}
\caption{\label{xe1272n} Cross-sections comparison for the production of $^{127}$Xe 
through $^{129}$Xe(n,3n)$^{127}$Xe 
between Geant4, ACTIVIA, and TENDL-2014. 
}
\end{figure}
\begin{figure}
\includegraphics[width = \textwidth]{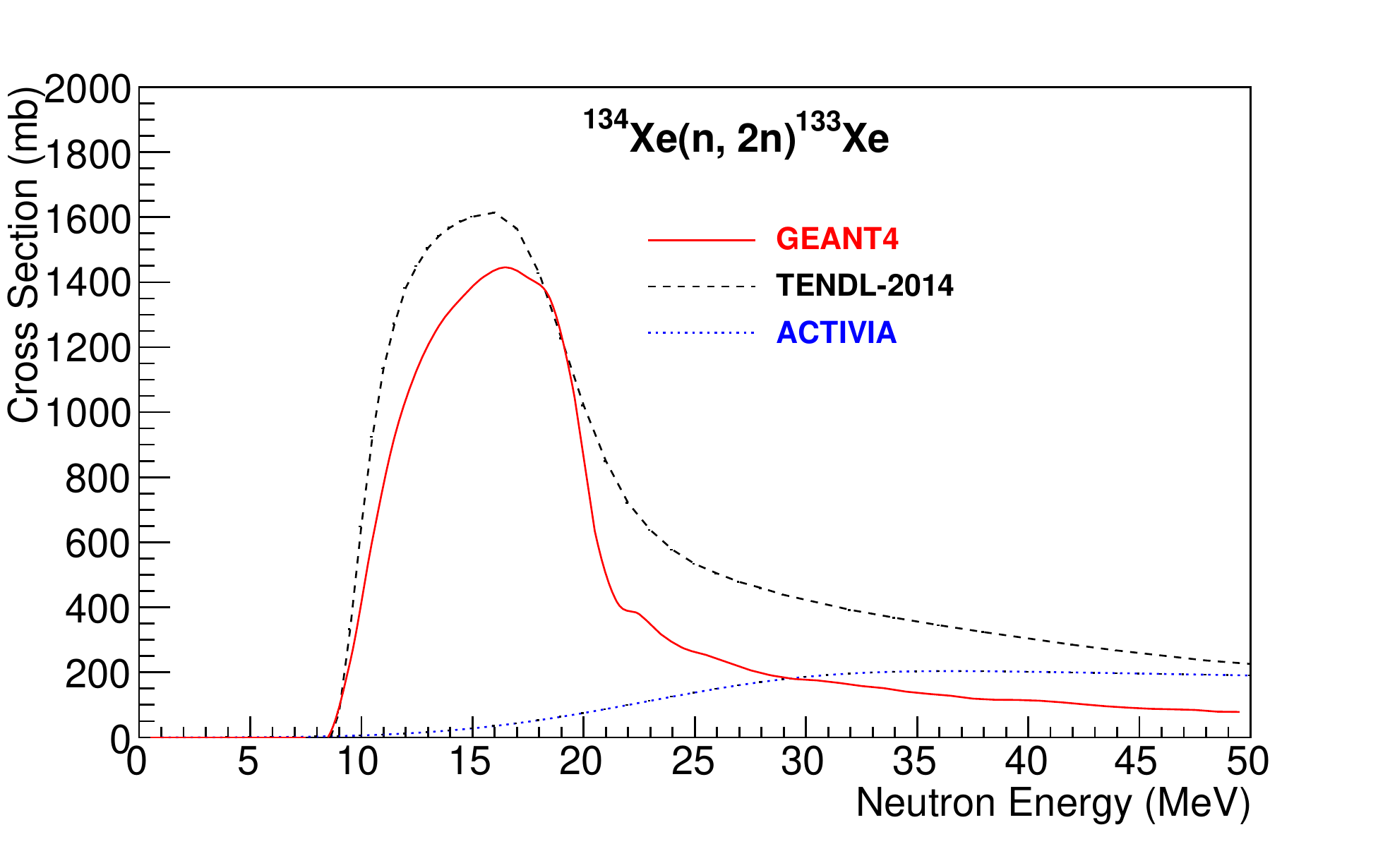}
\caption{\label{xe133} Cross-sections comparison for the production of $^{133}$Xe 
through $^{134}$Xe(n,2n)$^{133}$Xe
between Geant4, ACTIVIA, and TENDL-2014. 
}
\end{figure}

In the Geant4-based simulation, we track the initial and secondary particles to estimate the cosmogenic production rates. 
For a
given input spectrum of muons and neutrons, using a Geant4-based simulation and an ACTIVIA package, 
 the cosmogenically activated isotopes and their corresponding production rates are shown in Table~\ref{gasXenon}.
 Results presented in columns 3 and 6 (the 2nd value) were obtained with the same neutron spectrum from ``NY data'' above 4 MeV.
The difference in the production rates is then mainly due to the difference in the cross section libraries used in the Geant4-based simulation and ACTIVIA package and due to missing neutron 
transport and thermal neutron capture in ACTIVIA.
Note that the difference in the production rates due to the lack of tracking capability,
which takes into account the scattering/back-scattering processes, in ACTIVIA is expected to be small
since the simulated targets are small.
 We illustrate some of these differences 
in the cross sections for $^{128}$Xe($n$,2$n$)$^{127}$Xe, $^{129}$Xe($n$,3$n$)$^{127}$Xe, $^{134}$Xe($n$,2$n$)$^{133}$Xe 
in Figures~\ref{xe1271n},~\ref{xe1272n},~\ref{xe133}.
By default ACTIVIA uses the semi-empirical cross sections 
from Silberberg and Tsao~\cite{silb,tsao} but the possibility to use a different set of cross sections from 
MENDL libraries~\cite{yun} is foreseen although these cross sections are not provided with ACTIVIA. We have
not used MENDL libraries in this work.
\begin{table}[htbp]
\caption{\label{gasXenon}
Cosmogenic production rates in xenon gas at the sea level. Columns 2 - 5 are obtained from the Geant4 simulation.}
\begin{tabular}{|l|llll|l||c|}\hline
\multirow{2}{*}{Isotope, Half Life} & $n$($<$4MeV) &$n$($>$4MeV) & Muon & Proton & Total &ACTIVIA1/2\\
        & \multicolumn{4}{c|}{(kg$^{-1}$day$^{-1}$)}&(kg$^{-1}$day$^{-1}$)&(kg$^{-1}$day$^{-1}$)\\
\hline
$^{3}_{1}H$, 12.3y & 0 & 28.92 & 0.59 & 2.08 & 31.58 & 31.14/35.63\\
$^{81}_{36}Kr$, 2.3$\times$10$^{5}$y & 0 & 0.046 & 0 & 0.02 & 0.06& 0.42/0.20\\
$^{109}_{48}Cd$, 462.6d & 0 & 1.37 & 0 & 0.11 & 1.48 & 3.08/3.43\\
$^{113}_{50}Sn$, 115.1d & 0 & 6.50 & 0 & 0.40 & 6.90 & 4.39/5.89\\
$^{119}_{50}Sn$, 293.1d & 0 & 1.49 & 0 & 0.05 & 1.53 & 0.064/0.11\\
$^{124}_{51}Sb$, 60.2d & 0 & 1.59 & 0 & 0.04 & 1.62 &0.030/0.017\\
$^{125}_{51}Sb$, 2.8y & 0 & 1.45 & 0 & 0.03 & 1.48&0.016/0.009\\
$^{121}_{52}Te$, 154d & 0 & 20.60 & 0.099 & 0.50 & 21.20 &25.88/54.46\\
$^{123}_{52}Te$, 119.7d & 0 & 17.93 & 0.20 & 0.35 & 18.47 &1.27/2.67\\
$^{125}_{53}I$, 59.4d & 0 & 76.14 & 0.39 & 1.08 & 77.61 &37.35/88.67\\
$^{129}_{53}I$, 1.57$\times$10$^{7}$y & 0 & 76.23 & 0.30 & 0.82 & 77.35 &28.53/77.23\\
$^{127}_{54}Xe$, 36.4d & 0.64 & 228.8 & 1.48 & 2.42 & 233.30 &35.72/89.94\\
$^{133}_{54}Xe$, 5.2d & 11.08 & 85.58 & 1.68 & 0.85 & 99.19 &13.76/33.63\\
$^{135}_{54}Xe$, 9.1h & 2.18 & 56.19 & 1.09 & 0.60 & 60.05 &6.79/16.57\\
\hline
\end{tabular}
\end{table}

TENDL is a nuclear data library used in the TALYS nuclear physics code. 
Geant4 uses a combination of the evaluated data libraries (ENDF/B$-$VII.0~\cite{fkm}, 
JEFF$-$3.1~\cite{asa}, JENDL$-$4.0~\cite{ksh}, etc). 
As can be seen in Figures~\ref{xe1271n},~\ref{xe1272n},~\ref{xe133}, the cross sections
from Genat4 are similar to the cross sections from TENDL. The cross sections from
ACTIVIA are significantly different from Genat4 and TENDL.

\subsection{Cosmogenic production rates in some key components of the rare event physics experiments}
Other than the xenon target itself, key materials, such as polytetrafluoroethylene (PTFE), copper,
titanium, and stainless steel, which 
are generally used to build a detector, can also contribute to radioactivity. In the simulations,
a bulk cylinder of these materials is assumed with 10 cm in diameter and 
10 cm in height for evaluation. 
\par  
PTFE 
is commonly used as an excellent reflector in experiments detecting light from scintillations, for instance 
in liquid xenon detectors. Cosmic rays can produce radioactive isotopes 
such as $^{7}$Be, $^{10}$Be and $^{14}$C in PTFE. The simulated production rates are listed 
in Table~\ref{Teflon}. The difference in the production rates between the Geant4 simulation
and ACTIVIA2 caused by cross sections ranges from a factor of 4 to 3000. However, the production
 difference 
due to the difference in neutron flux between ACTIVIA1 and ACTIVIA2 is only about a factor of 2. 
\begin{table}[htbp]
\caption{\label{Teflon}
Cosmogenic production in PTFE at the sea level. Columns 2 - 5 give Geant4 results. }
\begin{tabular}{|l|llll|l||c|}\hline
\multirow{2}{*}{Isotope,Half Life} & n($<$4MeV) &n($>$4MeV) & Muon & proton & Total & ACTIVIA1/2\\
        & \multicolumn{4}{c|}{(kg$^{-1}$day$^{-1}$)}&(kg$^{-1}$day$^{-1}$)& (kg$^{-1}$day$^{-1}$)\\
\hline
$^{7}_{4}Be$, 53.1d  & 0.00 & 15.80 & 0.04 & 0.96 & 16.81 & 27.88/60.81 \\
$^{10}_{4}Be$, 1.5$\times$10$^{6}$y & 0.00 & 65.34 & 0.05 & 0.95 & 66.35 & 4.99/9.01 \\
$^{14}_{6}C$, 5.7$\times$10$^{3}$y & 0.09 & 0.01 & 0.00 & 0.00 & 0.10& 13.74/29.62 \\
\hline
\end{tabular}
\end{table}

\par
\begin{figure}
\includegraphics[width = \textwidth]{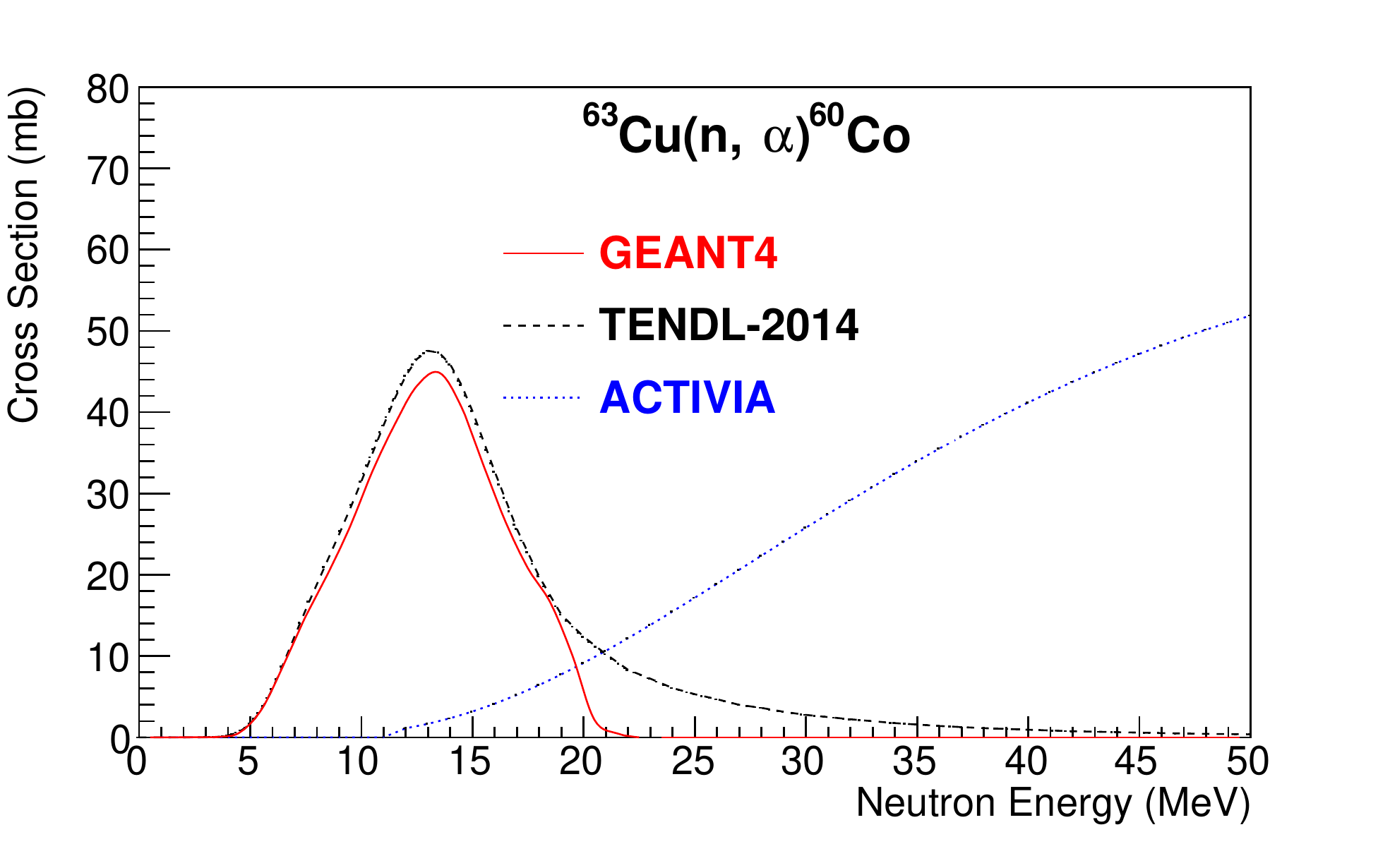}
\caption{\label{co60} The cross-section comparison for the production of $^{60}$Co 
through $^{63}$Cu($n$,$\alpha$)$^{60}$Co 
between Geant4, ACTIVIA, and TENDL-2014. 
}
\end{figure}

Oxygen-free high-conductivity copper (OFHC) or electroformed copper can be made highly radiopure.
 The radiopure copper is commonly used as  
shielding and structural material for low-background detectors. The 
concern for this material when it is produced at the surface is the cosmogenically produced isotopes such as 
$^{60}$Co, $^{57}$Co, $^{54}$Mn.
The cosmogenic production rates for copper are shown in 
Table~\ref{Copper}. The difference in the production rates
between the Geant4 simulation and ACTIVIA2 caused by cross sections ranges from a factor of 1.05 ($^{60}$Co)
to 9 ($^{44}$Ti). A good agreement between the Geant4 simulation and 
ACTIVIA2 for the production
of $^{60}$Co shown in Table~\ref{Copper} is not due to the agreement in cross sections used
in the two packages. In fact, the cross sections for the main production
channel, $^{63}$Cu($n$,$\alpha$)$^{60}$Co, in Geant4 and ACTIVIA are very different as can be seen
in Figure~\ref{co60}. The difference between the cross-sections is then compensated
 by other effects such as the energy dependence of the cross-section and neutron flux, 
leading to an apparent similarity in production rates. Large cross-section in ACTIVIA is achieved 
at higher energies where the neutron flux is smaller,  but extends to much higher neutron energies 
than the Geant4 cross-section. Thus the convolution of the cross-section with the neutron flux done 
numerically (ACTIVIA) or by Monte Carlo (Geant4) may lead to very similar results.
 
 The difference in the production rates due to the different fluxes
between ACTIVIA1 and 2 is within a factor of 2, but not very meaningful given the previous remark.
Note that $^{60}$Co, produced through $^{63}$Cu($n$,$\alpha$)$^{60}$Co, is a main concern for both direct 
detection of dark matter and
neutrinoless double-beta decay experiments because it has a long half-life (5.3 years) and emits
two gamma rays (1.173 MeV and 1.333 MeV) with a summed energy up to 2.506 MeV. These two gamma
rays undergoing Compton scattering can generate background events in the region of interest for
dark matter experiments. Two gamma rays can summed up to 2.506 MeV energy, which is in the
region of interest for neutrinoless double-beta decay experiments.      

\begin{table}[htbp]
\caption{\label{Copper}
Cosmogenic production in copper at the sea level. Columns 2 - 5 give Geant4 results.}
\begin{tabular}{|l|llll|l||c|}\hline
\multirow{2}{*}{Isotope,Half Life} & n($<$4MeV) &n($>$4MeV) & Muon & proton & Total & ACTIVIA1/2\\
        & \multicolumn{4}{c|}{(kg$^{-1}$day$^{-1}$)}&(kg$^{-1}$day$^{-1}$)& (kg$^{-1}$day$^{-1}$)\\
\hline
$^{22}_{11}Na$, 2.6y & 0 & 0.012 & 0.0027 & 0.002 & 0.014& 0.31/0.19\\
$^{26}_{13}Al$, 7.2$\times$10$^{5}$y & 0 & 0.016 & 0.0027 & 0.002 & 0.021& 0.23/0.14\\
$^{32}_{14}Si$, 150y & 0 & 0.063 & 0.0027 & 0.002 & 0.068 &  0.13/0.092\\
$^{40}_{19}K$, 1.3$\times$10$^{9}$y & 0 & 0.48 & 0.022 & 0.04  & 0.54 & 1.82/1.75\\
$^{47}_{20}Ca$, 4.5d & 0 & 0.12 & 0.0 & 0.01 & 0.13 &  0.026/0.036\\
$^{46}_{21}Sc$, 83.8d & 0 & 1.05 & 0.024 & 0.12 & 1.19 & 3.13/4.09\\
$^{47}_{21}Sc$, 3.3d & 0 & 0.98 & 0.011 & 0.10 & 1.09 & 0.62/0.86\\
$^{44}_{22}Ti$, 63y & 0 & 1.72 & 0.05 & 0.26 & 2.02 & 0.16/0.19\\
$^{50}_{23}V$, 1.4$\times$10$^{17}$y & 0 & 3.32 & 0.03 & 0.26 & 3.60 & 4.43/7.43\\
$^{51}_{24}Cr$, 27.7d & 0 & 15.20 & 0.12 & 1.16 & 16.48 &  10.00/18.08\\
$^{54}_{25}Mn$, 312.3d & 0 & 11.68 & 0.08 & 0.55 & 12.31 & 14.32/30.00\\
$^{55}_{26}Fe$, 2.7y & 0 & 53.66 & 0.25 & 2.43 & 56.33 & 19.32/42.79\\
$^{59}_{26}Fe$, 44.5d & 0 & 8.56 & 0.04 &  0.18 & 8.77 & 4.24/10.49\\
$^{60}_{26}Fe$, 1.5$\times$10$^{6}$y & 0.00 & 4.90 & 0.03 & 0.10 & 5.03 & 0.80/1.98\\
$^{56}_{27}Co$, 77.3d & 0 & 9.71 & 0.08 & 0.54 & 10.32 & 8.74/20.13\\
$^{57}_{27}Co$, 271.8d & 0 & 64.33 & 0.26 & 2.55 & 67.15 & 32.44/77.45\\
$^{58}_{27}Co$, 70.9d & 0 & 55.52 & 0.17 & 1.57 & 57.26 & 56.61/138.06\\
$^{60}_{27}Co$, 5.3y & 0.02 & 63.12 & 0.24 & 1.25 & 64.63 & 26.28/66.12 \\
$^{65}_{30}Zn$, 244.3d & 0 & 1.80 & 0.02 & 0.22 & 2.04 & 19.58/62.78 \\
\hline
\end{tabular}
\end{table}

\par
\begin{figure}
\includegraphics[width = \textwidth]{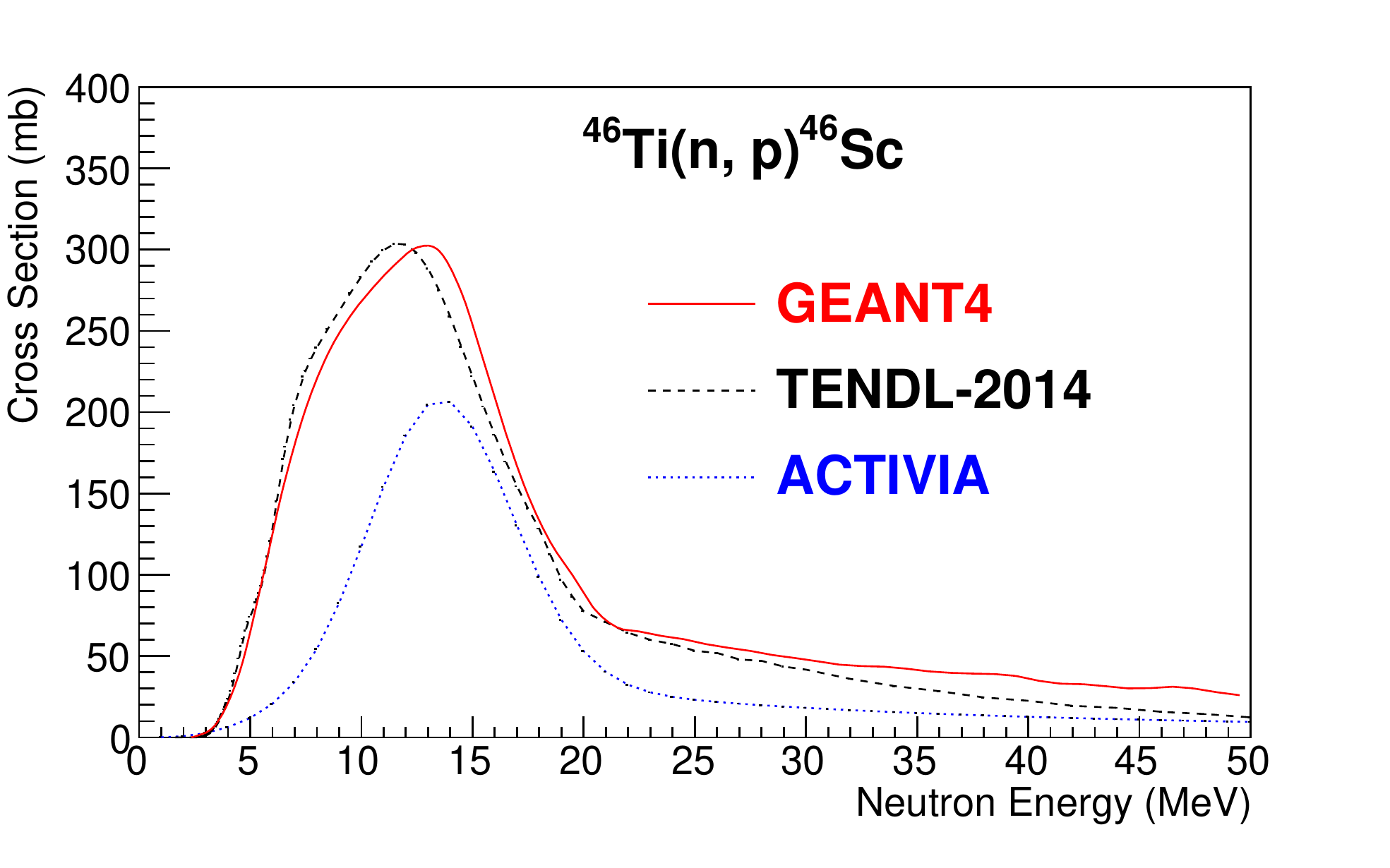}
\caption{\label{46sc} Cross-sections comparison for the production of $^{46}$Sc 
through $^{46}$Ti(n,p)$^{46}$Sc 
between Geant4, ACTIVIA, and TENDL-2014. 
} 
\end{figure}

Stainless steel is normally used to build detector vessels, or as a structural material
 for low-background experiments~\cite{xenon100, xmass}. Titanium was first adopted as a vessel material for 
liquid xenon by the LUX experiment~\cite{lux}
because a very radiopure sample has been found.  
 The simulated cosmogenic production rates for 
stainless steel and titanium can be found
in Table~\ref{Titanium} and~\ref{SSteel}. As can be seen in Table~~\ref{Titanium}, there is a good agreement
for the production rate of $^{46}$Sc between the Geant4 simulation and ACTIVIA2 with the same neutron energy
spectrum, although as can be seen in Figure~\ref{46sc}, the cross sections are different by about a factor 2. 
A factor of about 2 caused by the cross section difference is seen in 
the production of $^{46}$Sc in stainless steel 
between the Geant4 simulation and ACTIVIA2 (Table~\ref{SSteel}). 
Note that $^{46}$Sc with a half-life of 83.8 days is a concern to dark matter experiments because it produces two
simultaneous gamma rays with energies 0.889 MeV and 1.121 MeV that can undergo Compton scattering
to generate background events in the region of interest. The production of $^{46}$Sc in titanium
is through $^{46}$Ti($n$,p)$^{46}$Sc and in stainless steel is via spallation processes. 
\begin{table}[htbp]
\caption{\label{Titanium}
Cosmogenic production in titanium at the sea level. Columns 2 - 5 give Geant4 results.}
\begin{tabular}{|l|llll|l||c|}\hline
\multirow{2}{*}{Isotope,Half Life} & n($<$4MeV) &n($>$4MeV) & Muon & proton & Total & ACTIVIA1/2\\
        & \multicolumn{4}{c|}{(kg$^{-1}$day$^{-1}$)}&(kg$^{-1}$day$^{-1}$)& (kg$^{-1}$day$^{-1}$)\\
\hline
$^{22}_{11}Na$, 2.6y & 0.00 & 0.20 & 0.00 & 0.08 & 0.28 & 0.93/0.79 \\
$^{26}_{13}Al$, 7.2$\times$10$^{5}$y & 0.00 & 0.41 & 0.02 & 0.10 & 0.52 & 1.02/1.02 \\
$^{32}_{14}Si$, 150y & 0.00 & 1.51 & 0.00 & 0.14 & 1.65& 0.87/1.21 \\
$^{40}_{19}K$, 1.3$\times$10$^{9}$y & 0.00 & 20.98 & 0.12 & 0.97 & 22.06& 27.46/60.98 \\
$^{47}_{20}Ca$, 4.5d & 0.00 & 10.13 & 0.01 & 0.09 & 10.23& 0.31/0.72 \\
$^{46}_{21}Sc$, 83.8d & 0.31 & 270.42 & 0.43 & 4.32& 275.49& 107.80/270.07 \\
$^{47}_{21}Sc$, 3.3d  & 1.83 & 385.35 & 0.54 & 6.39 & 394.12& 48.78/116.38 \\
$^{44}_{22}Ti$, 63y & 0.00 & 95.94 & 0.46 & 3.68 & 100.08 & 4.84/12.03 \\
$^{50}_{23}V$, 1.4$\times$10$^{7}$y & 0.00 & 0.32 & 0.01 & 0.07 & 0.39& 4.40/14.13 \\
$^{51}_{24}Cr$, 27.7d & 0.00 & 0.02 & 0.00 & 0.00& 0.02& 0/0 \\
\hline
\end{tabular}
\end{table}

\begin{table}[htbp]
\caption{\label{SSteel}
Cosmogenic production in stainless steel at the sea level. Columns 2 - 5 give Geant4 results.}
\begin{tabular}{|l|llll|l||c|}\hline
\multirow{2}{*}{Isotope,Half Life} & n($<$4MeV) &n($>$4MeV) & Muon & proton & Total & ACTIVIA1/2\\
        & \multicolumn{4}{c|}{(kg$^{-1}$day$^{-1}$)}&(kg$^{-1}$day$^{-1}$)& (kg$^{-1}$day$^{-1}$)\\
\hline
$^{7}_{4}Be$, 53.1d & 0.00 & 0.02 & 0.01 & 0.02 & 0.05 & 2.04/2.05 \\
$^{10}_{4}Be$, 1.5$\times$10$^{6}$y & 0.00 & 0.10 & 0.00 & 0.01 & 0.11 & 0.93/0.89 \\
$^{14}_{6}C$, 5.7$\times$10$^{3}$y & 0.60 & 1.16 & 0.01 & 0.04 & 1.81 & 0.41/0.28 \\
$^{22}_{11}Na$, 2.6y & 0.00 & 0.22 & 0.00 & 0.05 & 0.27 & 0.65/0.69 \\
$^{26}_{13}Al$, 7.2$\times$10$^{5}$y & 0.00 & 0.80 & 0.01 & 0.07 & 0.88& 0.97/1.57 \\ 
$^{32}_{14}Si$, 150y & 0.00 & 0.32 & 0.02 & 0.05 & 0.39 & 0.31/0.30 \\ 
$^{40}_{19}K$, 1.3$\times$10$^{9}$y  & 0.00 & 2.60 & 0.04 & 0.26 & 2.90 & 5.94/8.90 \\
$^{47}_{20}Ca$, 4.5d & 0.00 & 0.97 & 0.00 & 0.03 & 1.00 & 0.18/0.42 \\
$^{46}_{21}Sc$, 83.8d & 0.00 & 8.43 & 0.04 & 0.34& 8.80 & 8.09/17.84 \\
$^{47}_{21}Sc$, 3.3d & 0.00 & 8.77 & 0.02 & 0.29 & 9.08 & 3.54/8.14 \\
$^{44}_{22}Ti$, 63y & 0.00 & 12.14 & 0.15 & 0.98 & 13.27 & 0.86/1.69 \\
$^{50}_{23}V$, 1.4$\times$10$^{17}$y  & 0.25 & 70.58 & 0.24 & 1.55 & 72.62 & 42.06/102.84 \\
$^{51}_{24}Cr$, 27.7d & 3.94 & 282.47 & 5.34 & 6.86 & 298.61 & 88.92/222.34 \\
$^{54}_{25}Mn$, 312.3d & 3.53 & 222.24 & 0.69 & 4.00 & 230.45 & 74.75/191.02 \\
$^{55}_{26}Fe$, 2.7y & 3.78 & 621.84 & 18.30 & 12.99 & 656.90 & 106.45/266.52 \\
$^{59}_{26}Fe$, 44.5d & 0.14 & 0.31 & 0.00 & 0.01 & 0.45& 0.08/0.20 \\
$^{60}_{26}Fe$, 1.5$\times$10$^{6}$y & 0.00 & 0.08 & 0.00 & 0.00 & 0.08& 0.02/0.05 \\
$^{56}_{27}Co$, 77.3d & 0.00 & 14.68 & 0.10 & 0.97 & 15.75 & 47.71/130.59 \\
$^{57}_{27}Co$, 271.8d & 0.00  & 79.12 & 0.15 & 1.47 & 80.74 & 14.74/36.07 \\
$^{58}_{27}Co$, 70.9d & 8.15 & 80.07 & 0.39 & 1.51 & 90.11 & 4.95/13.04 \\
$^{60}_{27}Co$, 5.3y & 0.00 & 6.17 & 0.01 & 0.10& 6.27 & 1.81/4.92 \\
\hline
\end{tabular}
\end{table}

\par
\section{Comparison between calculations and the available experimental data}
As can be seen from Tables 1-5, the results from the calculations using ACTIVIA 
(with default cross sections from~\cite{silb, tsao})
are inconsistent with
the simulated results from Geant4 for many isotopes. The inconsistencies can be caused by
 (1) the neutron energy spectrum and (2)
the library for cross sections of inelastic scattering processes. To understand which simulation tool 
delivers more reliable results, we compared the Geant4 simulation results with
ACTIVIA as well as the available experimental data from LUX~\cite{luxback} and Baudis et al.~\cite{xenon100back, mla}.

The cosmogenic production rates from the Geant4-based simulations and the calculations of ACTIVIA are
converted into radioactivity (decay rate) using the formula below:
\begin{equation}
A(Bq/kg)= \frac{(R\times(1-e^{(-ln2\times T_{s}/t_{1/2})}))\times e^{(-ln2\times T_{u}/t_{1/2})}}{86400},
\end{equation}
where $R$ is the cosmogenic production rate in atoms per kg per day, $T_{s}$ is the exposure time on the surface,
$t_{1/2}$ is the decay half-life, 
and $T_{u}$ is the decay time at the underground site before the measurement is taken. Note that
the production rates in the simulations need to be corrected by altitudes where the experimental
data were taken. The fluxes of muon and neutrons as a function of altitude is shown in
Figure~\ref{altitude}.  
\subsection{Comparison for the cosmogenic activity in natural xenon}
The LUX~\cite{lux} experiment reported the measured activity of $^{127}$Xe and $^{133}$Xe 
utilizing their first three months
of data~\cite{luxback, lux}. In this work, an exposure time period of 150 days at sea level 
and appropriate exposures (7 to 49 days) at an altitude of 1480 meters 
are applied according to information recorded historically
since the xenon gas bottles were produced~\cite{luxback}. Predicted and observed decay rates are
listed in Table 6 after 90 days underground.
\begin{table}[htbp]
\caption{\label{Comparison}
Decay rates of two xenon isotopes as calculated with Geant4 and ACTIVIA, and measured by LUX~\cite{luxback, lux}
after 90 days cooling down underground.
}
\begin{tabular}{|l|l|c|c|c|}\hline
Activated & Target & This Work &  ACTIVIA1/2 & LUX data\cite{luxback}\\
Isotope   &        & ($\mu$Bq/kg) & ($\mu$Bq/kg) & ($\mu$Bq/kg) \\
\hline
$^{127}Xe$ & Xe & 470 & 73/180 & (490$\pm$95) \\
$^{133}Xe$ & Xe & $7.0\times10^{-3}$ & $9.8\times10^{-4}/2.4\times10^{-3}$ & $(25.0\pm5.0)\times10^{-3}$ \\
\hline
\end{tabular}
\end{table}
\par
The activation of xenon is also compared to the available data from Baudis et al.~\cite{xenon100back}
as shown in Table~\ref{comxenon1001} in where the saturation activity at sea level is used. Note
that the saturation activity describes a saturation level at which its disintegration rate equals
its production rate and allows a fair comparison between simulations/calculations and experimental data.
\begin{table}[htbp]
\caption{\label{comxenon1001}
Saturation activity at sea level assuming infinite exposure time.
}
\begin{tabular}{|l|l|c|c|c|c|}\hline
Activated & Target & This Work &  ACTIVIA1/2 & Baudis et al. data~\cite{xenon100back}\\
Isotope   &        & ($\mu$Bq/kg) & ($\mu$Bq/kg) & ($\mu$Bq/kg) \\
\hline
$^{113}$Sn & Xe & 67.6 &45.7/61.3& $<$55      \\
$^{125}$Sb & Xe & 16.8&0.2/0.01& 590$^{+260}_{-230}$\\
$^{127}$Xe & Xe & 2670 & 413/1040&1870$^{+290}_{-270}$ \\
$^{133}$Xe & Xe & 1140 & 160/390& $<$1200 \\
\hline
\end{tabular}
\end{table}
\par
It is clear that the results for $^{127}$Xe and $^{133}$Xe from the Geant4 simulation agree 
with the results from LUX and Baudis et al. reasonably well. However, 
the results from ACTIVIA are different from LUX and Baudis et al. 
by a factor of more than 2 depending on the neutron flux used in
ACTIVIA. For example, for the production of $^{127}$Xe, the difference between ACTIVIA1 and the LUX data is
a factor of 6.7 and the difference between ACTIVIA2 and the LUX data is a factor of 2.7 with 
ACTIVIA2 calculations that used more accurate neutron flux.
This indicates a difference of a factor of 2.5 between ACTIVIA1 and 2 due 
to the difference in the assumed neutron flux.
The difference of a factor of 2.7 between ACTIVIA2 and the LUX data may come from the cross section libraries used in the calculation
of ACTIVIA2. 
We can conclude that the libraries for the cross sections
of neutron inelastic scattering in Geant4, at least, for the production of $^{127}$Xe and $^{133}$Xe
are adequate. 
\par
\subsection{Comparison for the cosmogenic activity in natural titanium}
Cosmogenic activation of natural titanium can produce a radioactive isotope of $^{46}$Sc. We compare the results from the Geant4
simulation and ACTIVIA calculations to the LUX data~\cite{luxback} in Table~\ref{titan}, using an exposure time
of 180 days 
for the titanium at 1480 meters altitude roughly equivalent to the LUX titanium history. 
As can be seen in Table~\ref{titan}, both
Geant4 simulation and ACTVIA agree with the measured data reasonably well ($<$a factor of 2 difference). 
This indicates that the cross section
libraries used in Geant4 and ACTIVIA are adequate for the production of $^{46}$Sc. 
\begin{table}[htbp]
\caption{\label{titan}
Saturation activity of $^{46}$Sc from  titanium activation at sea level.
}
\begin{tabular}{|l|l|c|c|c|}\hline
Activated & Target & This Work &  ACTIVIA1/2 & LUX data\cite{luxback}\\
Isotope   &        & ($\mu$Bq/kg) & ($\mu$Bq/kg) & ($\mu$Bq/kg) \\
\hline
$^{46}$Sc & Ti & 7300& 2900/7300 & (4400$\pm$300) \\
\hline
\end{tabular}
\end{table} 
\subsection{Comparison for the cosmogenic activity in natural copper}
Cosmogenic activation of natural copper 
from a Geant4 simulation are compared to ACTIVIA, Cosmo~\cite{xenon100back}, TALYS~\cite{mei}, 
and the available data~\cite{xenon100back} as shown
in Table~\ref{comxenon100}. The agreements between Geant4 and other calculations as well as the measurements from
Baudis et al. are within a factor of 2.  One must point out that even though 
 ACTIVIA1 with the original neutron spectrum (blue line in Fig. 1) 
shows better agreement with data than ACTIVIA2, this is not a valid
agreement as discussed previously in section 2.2 and Figure~\ref{co60} because 
the neutron spectrum used in 
ACTIVIA2 is more accurate based on the fit to the data.
\begin{table}[htbp]
\caption{\label{comxenon100}
Saturation activity of different isotopes from activation of copper at sea level.
}
\begin{tabular}{|l|l|c|c|c|c|}\hline
Activated & Geant4 &  ACTIVIA1/2 & Cosmo & TALYS~\cite{mei}& Baudis et al. data~\cite{xenon100back}\\
Isotope   & ($\mu$Bq/kg) & ($\mu$Bq/kg) & ($\mu$Bq/kg) & ($\mu$Bq/kg) & ($\mu$Bq/kg)\\
\hline
$^{46}$Sc &12.4 &36.2/47.4 & 17.0 & - &  27$^{+11}_{-9}$     \\
$^{54}$Mn &136.0 &165.7/347.2 &156 & 188&  154$^{+35}_{-34}$     \\
$^{59}$Fe &99.5 &49.1/121.5 &50 & -&    47$^{+16}_{-14}$   \\
$^{56}$Co &113.2 &101.2/233 &81 & -&    108$^{+14}_{-16}$   \\
$^{57}$Co &747.7 & 375.5/896.4&350 &650 &  519$^{100}_{-95}$     \\
$^{58}$Co & 644.4& 655.2/1597.8&632 & -&  798$^{+62}_{-58}$     \\
$^{60}$Co & 713.2& 295.7/744&297 &537 &   340$^{+82}_{-68}$    \\
\hline
\end{tabular}
\end{table}
\subsection{Tritium production}
$^{3}$H, a long lived radioactive isotope ($\beta$ decay), 
can be produced in all materials through muon spallation and neutron inelastic scattering. 
In Table~\ref{H3} we tabulate the cosmogenic production rate of $^{3}$H in various targets from the Geant4 simulation and other available calculations. It is worth mentioning that
the production of $^{3}$H in germanium and NaI targets from Geant4 
is slightly higher than the earlier calculation using TALYS 1.0~\cite{mei}. 
A total of 42.87 kg$^{-1}$d$^{-1}$ in NaI is consistent with a recent report 
of $\sim$40.6 kg$^{-1}$d$^{-1}$~\cite{jam} from ANAIS-25 as
shown in Table~\ref{H3}. Aside from silicon, the agreement is good between the Geant4 simulation
and ACTIVIA2. 
\begin{table}[htbp]
\caption{\label{H3}
The cosmogenic production rate of $^{3}H$ at the sea level in various targets. Columns 2 -5 
are obtained from the Geant4 simulation. }
\begin{tabular}{|l|lll|c|c|c|c|}\hline
\multirow{2}{*}{Target} & $n$($<$4 MeV) &$n$($>$4 MeV) & $\mu$ & Total&ACTIVIA1/2 &TALYS& ANAIS-25\\
        & \multicolumn{3}{c|}{(kg$^{-1}$day$^{-1}$)}&(kg$^{-1}$day$^{-1}$) &(kg$^{-1}$day$^{-1}$)&(kg$^{-1}$day$^{-1}$)&(kg$^{-1}$day$^{-1}$)\\
\hline
C$_{2}$H$_{6}$ & 0 & 279.1 & 0.374 & 279.5&-/-&- &- \\
Si& 0 & 27.12 & 0.1668 & 27.29 &56.54/108.74&- & -\\
Argon& 0 & 84.49 & 0.4135 & 84.91&45.58/82.9&44.4 &- \\
Ge& 0 & 48.21 & 0.1127 & 48.32 &34.13/52.39 &27.7&-\\
NaI& 0 & 42.6 & 0.2629 & 42.87 &31.9/36.19&31.1 &40.6\\
Xenon& 0 & 32.13 & 0.6348 & 32.76 &31.14/35.63&16.0 &- \\
\hline
\end{tabular}
\end{table}
\section{Cosmogenic and radiogenic production underground}
Cosmogenic activation of materials in a underground environment can be estimated using the method described
in Barker et al.~\cite{barker}. Muon and muon-induced neutron fluxes 
are several orders of
magnitude lower underground than at surface. The cosmogenic production underground
is therefore often expected to be negligible. Since the radioactive isotopes in Tables 1-5 are mainly induced
by fast neutrons from cosmic rays, and there are fast neutrons from natural radioactivity
in any underground laboratory, we also evaluate radiogenic production for the most important
isotopes relevant to xenon-based experiments.  
From the results of the simulations in Tables 1-5, the isotopes with higher production rates, longer half-life,
and emission of either low energy X-rays/Auger electrons ($^{127}$Xe) inside target or gamma rays with energies up to MeV
($^{46}$Sc and $^{60}$Co) in titanium or copper that are close to target, are problematic. Thus,  
the main concerns are the production
 of $^{127}$Xe in xenon gas, $^{46}$Sc in titanium, and $^{60}$Co in copper and stainless steel. 
To understand the production rates underground, we use $^{127}$Xe as an example.
  The production rate is 229 atoms kg$^{-1}$d$^{-1}$
 on the surface with a fast neutron flux (above 4 MeV) of 0.004 neutrons cm$^{-2}$s$^{-1}$. 
 The capture process associated with thermal neutrons contributes 0.64 atoms kg$^{-1}$d$^{-1}$
 on the surface with a flux of 0.002 neutrons cm$^{-2}$s$^{-1}$. We simulated the production rate for an underground laboratory 
where the fast neutron flux (E$_{n}$ $>$ 4 MeV) is about 10$^{-6}$ neutrons cm$^{-2}$s$^{-1}$ 
and the thermal neutron flux is about 10$^{-7}$ neutrons cm$^{-2}$s$^{-1}$~\cite{mei1}, using Geant4
to estimate the production rate of radioactive isotopes underground. 
The neutron energy spectrum from radiogenic processes obtained in Ref.~\cite{mei1},
where U/Th contents and the density of rock are described, is used in the Geant4 simulation.
We summarize the production rate of $^{127}$Xe at SURF, 
$\sim$1480 meters below the surface~\cite{mei1, jaret}. Note that the calculation is performed for xenon gas bottles without 
any shielding in the
underground space.
\begin{itemize}
\item{The possible maximum production rate from fast neutrons is less than 
2.4$\times$10$^{-3}$ atoms kg$^{-1}$d$^{-1}$. The dominant activation rate is due to either ($\alpha$, n) or spontaneous fission neutrons (depending on the rock composition) with neutron energies above 4 MeV, resulting from natural radioactivity~\cite{vto}.} 
\item{The possible production rate from thermal neutrons is about 7.5$\times$10$^{-5}$ atoms kg$^{-1}$d$^{-1}$.} 
\item{The possible production rate from cosmogenic activation through spallation is estimated to be 7.1$\times$10$^{-4}$
 atoms kg$^{-1}$d$^{-1}$. For the neutrons from cosmic-ray muons, an approximate scaling method for neutron spectrum, 
which takes into account the average energy difference between the sea level and the underground, 
is used as described in Ref.~\cite{barker}. The scaling factor is ($\frac{E_{\mu}^{ug}}{E_{\mu}^{su}})^{0.73}\times\frac{\Phi_{\mu}(ug)}{\Phi_{\mu}(su)}\times R_{iso}(su)$, where E$_{\mu}^{ug}$ = 321 GeV is the average energy of muons at a depth of 1480 meters~\cite{barker}, E$_{\mu}^{su}$ = 4 GeV is the average energy of muons at the surface~\cite{barker}, $\Phi_{\mu}(ug)$ = 4.4$\times$10$^{-9}$ cm$^{-2}$s$^{-1}$ is the total muon flux at a depth of 1480 meters, $\Phi_{\mu}(su)$ = 0.0347 cm$^{-2}$s$^{-1}$ is the average muon flux weighted with the exposure dates at sea level and at the surface of SURF, R$_{iso}$ is the cosmogenic activation rate induced by neutrons at the surface.} 
\end{itemize}

The conclusion is that the production of $^{127}$Xe underground is about 3 atoms per ton per day with
a saturation activity of 3.5$\times$10$^{-5}$ Bq/ton. This is a very low activity 
and only a tiny fraction of decays through emission of low energy X-rays/Auger electrons can cause 
background events for xenon-based dark matter experiments. Moreover, 
such sensitive detectors are protected by neutron shielding so during
 the science run the background rate due to activation is much smaller 
than from radioactivity in detector components. Activation induced 
before the target is shielded (during underground storage) will 
decay within a few months during commissioning phase of the experiment.
We conclude that the radioactivity of 
these isotopes produced underground can be neglected.

\section{Conclusion}
The cosmogenic activations of several key components of the next generation 
rare event search experiments at sea level have been simulated using 
the GEANT4 and ACTIVIA packages. Fast neutrons, thermal neutrons and muons at the Earth surface 
are considered individually. 
 The total production rates of several isotopes are compared 
with other calculations~\cite{gerda, MJ, mei} as well as the available experimental data from LUX~\cite{luxback} and Baudis et al.~\cite{xenon100back}. 
Significant differences are found in the 
results between Geant4 and ACTIVIA packages. We attribute these differences to the different input 
neutron energy spectra and the cross-section libraries used in the two packages. 
We emphasize the key requirement of using the accurate neutron energy spectrum and cross-sections 
for particular nuclear reactions. We have found that Geant4 can predict the production rates of different 
radioactive isotopes by cosmogenic activation pretty well.

\section{Acknowledgments}
The authors wish to thank Christina Keller and Wenzhao Wei for carefully reading of this manuscript. 
This work is supported in part by NSF PHY-0758120, PHYS-0919278,
PHYS-1242640, DOE grant DE-FG02-10ER46709,
the Office of Research at the University
of South Dakota and a 2010 research center support by the State of South Dakota.
The simulations of this work was performed on High Performance Computing systems at
the University of South Dakota. Electronic data exchange for this project was supported 
in part by NSF award ACI-1440681. V. A. Kudryavtsev contribution was 
supported by the Science and Technology Facilities Council (UK).

\end{document}